\documentstyle[11pt,epsfig]{article}
\begin{document}

\title{{\bf Noether symmetric $f(R)$ quantum cosmology and its classical correlations}}
\author{ Babak Vakili\thanks{%
email: b-vakili@sbu.ac.ir, bvakili45@gmail.com} \\
%EndAName
{\small {\it Department of Physics, Azad University of Chalous,}}\\{\small {\it P. O. Box 46615-397, Chalous, Iran}}} \maketitle

\begin{abstract}
We quantize a flat FRW cosmology in the context of the $f(R)$ gravity by Noether symmetry approach. We explicitly
calculate the form of $f(R)$ for which such symmetries exist. It is shown that the existence of a Noether symmetry yields a general
solution of the Wheeler-DeWitt equation where can be expressed as a superposition of states of the form $e^{iS}$. In terms of Hartle criterion,
this type of wave function exhibits classical correlations, i.e. the emergent of classical universe is expected due
to the oscillating behavior of the solutions of Wheeler-DeWitt equation. According to this interpretation
we also provide the Noether symmetric classical solutions of our $f(R)$ cosmological model.
\vspace{5mm}\newline
\end{abstract}

\section{Introduction}
In recent years, modified theories of gravity constructed by
adding correction terms to the usual Einstein-Hilbert action, have
opened a new window to study the accelerated expansion of the
universe. It has been shown that such correction terms could
give rise to accelerating solutions of the field equations without
having to invoke concepts such as dark energy \cite{1}. In a more general setting, one can use a generic function $f(R)$,
instead of the usual Ricci scalar $R$ as the action of general relativity.
Such $f(R)$ gravity theories have been extensively studied in the
literature over the past few years, see \cite{TS} for a review.  In finding the dynamical
equations of motion one can vary the action with respect to the
metric (metric formalism), or view the metric and connections as
independent dynamical variables and vary the action with respect
to both independently (Palatini formalism) \cite{2}. In this
theory, the Palatini form of the action is shown to be equivalent
to a scalar-tensor type theory from which the scalar field
kinetic energy is absent. This is achieved by introducing a
conformal transformation in which the conformal factor is taken as
an auxiliary scalar field \cite{3}. As is well known, in the usual
Einstein-Hilbert action these two approaches give the same field
equations. However, in $f(R)$ gravity the Palatini formalism leads
to different dynamical equations due to nonlinear terms in the
action. There is also a third version of $f(R)$ gravity in which
the Lagrangian of the matter depends on the connections of the
metric (metric-affine formalism) \cite{4}.

In a previous work \cite{5}, we studied a flat FRW space-time in the framework of
the metric formalism of $f(R)$ gravity. In \cite{5} we
constructed an effective Lagrangian in the minisuperspace $\{a, R\}$ where
$a$ and $R$ being the scale factor and Ricci scalar respectively.
The form of the function $f(R)$ appearing in the
modified action is then found by demanding that the Lagrangian
admits the desired Noether symmetry \cite{6}. A similar study of this issue in the Palatini framework can be found in \cite{FS}. By the Noether
symmetry of a given minisuperspace cosmological model we mean that
there exists a vector field $X$, as the infinitesimal generator of
the symmetry on the tangent space of the configuration space such
that the Lie derivative of the Lagrangian with respect to this
vector field vanishes. For some applications of the Noether symmetry approach in
various cosmological models see \cite{7}.

As mentioned above, although the corrections to the results of standard general relativity are widely investigated in literature in the
$f(R)$ gravity context, these works are often in the classical regimes \cite{TC}. The cases dealing with quantum $f(R)$ models have seldom been studied in
the literature \cite{8}, and it would be of interest to employ such models in this study.

In this paper we consider the same model as in \cite{5} and try to quantize it by Noether symmetry approach. In general, the existence of a
symmetry results in appearing of constants of motion which are related to the existence of cyclic variables in the dynamics \cite{9}.
This is the key result in our quantization procedure. Indeed, in terms of the cyclic variables the conserved quantities are nothing but
the corresponding conjugate momenta. We shall see that applying the quantum version of these symmetries on the wave function of the universe
(which satisfies the Wheeler-DeWitt equation) yields an oscillatory behavior for the wave function in the direction of the symmetries. On the other hand, in the semiclassical approximation for the canonical quantum gravity, one can show that a wave function with classical correlations is a superposition of states of the form $e^{iS}$, i.e. with oscillatory behavior \cite{10}. With this interpretation scheme for the solutions of Wheeler-DeWitt equation, we also obtain the classical trajectory and show that the cosmological scale factor obeys a power law expansion.

It is to be noted that our presentation of quantum cosmology is applying quantum mechanics to a reduced dynamical system, the so-called minisuperspace. This is because that the fundamental equation of quantum cosmology, the Wheeler-DeWitt equation, is a differential equation on the infinite dimensional superspace and dealing with its solutions in such a space is not an easy task. Therefore, in order to solve this equation and obtain the wave function of the universe we can use an approximation method in which one truncates the infinite degrees of freedom to a finite dimensional submanifold called minisuperspace. In an alternative approach to deal with quantum effects in the early universe one can use the idea of {\it quantum field theory in curved space time} \cite{BD}. In this theory we assume that the matter fields are quantized while the gravitational field is classical and given by Einstein or any other modified gravity field equations. In this scheme quantum gravity is truncated at one-loop level and for computation of the one-loop corrections quantum fluctuations are imposed on a given classical background. Our aim in this letter is to study of quantization procedure in a Noether symmetric $f(R)$ gravity model in a reduced phase space framework, i.e. as mentioned above we first fix the background and then try to quantize it by Wheeler-DeWitt approach. In \cite{QU} the role of quantum effects to the background cosmology at one-loop level are investigated in $f(R)$ gravity framework.
\section{The Noether symmetric phase space of the model}
In this section we consider a spatially flat FRW cosmology within
the framework of $f(R)$ gravity. Since our goal is to study models
which exhibit Noether symmetry, we do not include any matter
contribution in the action. Let us start from the
action (we work in units where $c=\hbar=16\pi G=1$)
\begin{equation}\label{A}
{\cal S}=\int d^4x\sqrt{-g}f(R),
\end{equation}
where $R$ is the scalar curvature and $f(R)$ is an arbitrary
function of $R$. By varying the above action with respect to
metric we obtain the equation of motion as
\begin{equation}\label{B}
\frac{1}{2}g_{\mu \nu}f(R)-R_{\mu
\nu}f'(R)+\nabla_{\mu}\nabla_{\nu}f'(R)-g_{\mu \nu}\Box f'(R)=0,
\end{equation}
where a prime represents differentiation with respect to $R$. We
assume that the geometry of space-time is described by the flat
FRW metric which seems to be consistent with the present
cosmological observations
\begin{equation}\label{C}
ds^2=-dt^2+a^2(t)\sum_{i=1}^{3}(dx^i)^2.
\end{equation}
With this background geometry the field equations read
\begin{equation}\label{D}
2\dot{H}+3H^2=-\frac{1}{f'}\left[f'''\dot{R}^2+2H\dot{R}f''+f''\ddot{R}+\frac{1}{2}
(f-Rf')\right],\end{equation}
\begin{equation}\label{E}
H^2=\frac{1}{6f'}\left[(f'R-f)-6\dot{R}Hf''\right],
\end{equation}
where $H=\dot{a}/a$ is the Hubble parameter and a dot represents
differentiation with respect to $t$. To study the symmetries of
the minisuperspace under consideration, we need an effective point-like
Lagrangian for the model whose variation with respect to its
dynamical variables yields the correct equations of motion.
Following \cite{11}, we consider the action described above as
representing a dynamical system in which the scale factor $a$ and
scalar curvature $R$ play the role of independent dynamical
variables. In \cite{5,9,11} it is shown that such point-like Lagrangian takes the form
\begin{equation}\label{F}
{\cal L}(a,\dot{a},R,
\dot{R})=6\dot{a}^2af'+6\dot{a}\dot{R}a^2f''+a^3(f'R-f).
\end{equation}
The Hamiltonian corresponding to  Lagrangian (\ref{F}) can then be
written in terms of $a$, $\dot{a}$, $R$ and $\dot{R}$ as
\begin{equation}\label{I}
{\cal H}(a,\dot{a},R,
\dot{R})=6\dot{a}^2af'+6\dot{a}\dot{R}a^2f''-a^3(f'R-f).
\end{equation}
Therefore, our cosmological setting is equivalent to a dynamical
system where the phase space is spanned by
$\left\{a,R,p_a,p_R\right\}$ with Lagrangian (\ref{F}) describing
the dynamics with respect to time $t$. Now, it is easy to see that
variation of  Lagrangian (\ref{F}) with respect to $R$ gives the
well-known relation for the scalar curvature, while variation with respect to $a$ yields the field equation
(\ref{D}). Also, equation (\ref{E}) is nothing but the zero energy
condition ${\cal H}=0$ (Hamiltonian constraint).

As is well known, Noether symmetry approach is a powerful tool in
finding the solution to a given Lagrangian, including the one
presented above. In this approach, one is concerned with finding
the cyclic variables related to conserved quantities and
consequently reducing the dynamics of the system to a manageable
one. The investigation of Noether symmetry in the model presented
above is therefore the goal we shall pursue here. Following \cite{6}, we
define the Noether symmetry induced on the model by a vector field
$X$ on the tangent space $TQ=\left(a,R,\dot{a},\dot{R}\right)$ of
the configuration space $Q=\left(a,R\right)$ of Lagrangian
(\ref{F}) as
\begin{equation}\label{J}
X=\alpha \frac{\partial}{\partial a}+\beta
\frac{\partial}{\partial R}+\frac{d
\alpha}{dt}\frac{\partial}{\partial \dot{a}}+\frac{d
\beta}{dt}\frac{\partial}{\partial \dot{R}},
\end{equation}
such that the Lie derivative of the Lagrangian with respect to
this vector field vanishes
\begin{equation}\label{K}
L_X {\cal L}=0.
\end{equation}
In (\ref{J}), $\alpha$ and $\beta$ are some unknown functions of $a$ and $R$.
Now, it is easy to see that the constants of motion corresponding to such a
symmetry are \cite{5,9}
\begin{equation}\label{P}
Q=\alpha p_a+\beta p_R.
\end{equation}
In order to obtain the functions $\alpha$ and $\beta$ we use
equation (\ref{K}). In general this equation gives a quadratic
polynomial in terms of $\dot{a}$ and $\dot{R}$ with coefficients
being partial derivatives of $\alpha$ and $\beta$ with respect to
the configuration variables $a$ and $R$. Thus, the resulting
expression is identically equal to zero if and only if these
coefficients are zero. This leads to a system of partial
differential equations for $\alpha$ and $\beta$ \cite{5}-\cite{7}. For the model at hand, this system of differential equations has been investigated
carefully in \cite{5} and it has been shown that in the case where $f''\neq 0$, we have the following solutions
\begin{equation}\label{AL}
\alpha(a)=-\frac{1}{2}a^{-1},\hspace{.5cm}\beta(a,R)=Ra^{-2},
\end{equation}
and \footnote{A remark about this form of $f(R)$ function is that it is found only by demanding that the Lagrangian admits the Noether
symmetry. Although, we shall see that this form of $f(R)$ gravity yields a power law inflationary cosmology, (see \cite{GC} for more general $f(R)$ models which lead to the inflationary and late-time accelerating epochs) but in view of having the correct weak-field limit at Newtonian and post-Newtonian levels has not a desired form. The conditions under which a modified gravity model passes the local and astrophysical tests such as Newton law and solar system tests are investigated in \cite{HU}. In these works such $f(R)$ theories are studied which satisfy the conditions \[\lim_{R\rightarrow \infty} f(R)=\mbox{cons.},\hspace{.5cm}\lim _{R\rightarrow 0}f(R)=0,\] and shown that they pass Newton law,
stability of Earth-like gravitational solution, heavy mass for additional scalar degree of freedom, etc.. Equation (\ref{AM}) shows that our Noether symmetric model does not satisfy the above conditions and hence is not a viable theory with correct Newtonian and post-Newtonian limits. This is not surprising since it is well-known that a large class of $f(R)$ theories suffer from this issue \cite{CH}.}
\begin{equation}\label{AM}
f(R)=R^{\frac{3}{2}}.
\end{equation}
Substituting this results into equation (\ref{F}) we obtain the Lagrangian of the Noether symmetric model as
\begin{equation}\label{Q1}
{\cal L}=9\dot{a}^2aR^{1/2}+\frac{9}{2}\dot{a}\dot{R}a^2R^{-1/2}+\frac{1}{2}a^3R^{3/2}.\end{equation}
The momenta conjugate to variables $a$ and $R$ are
\begin{equation}\label{Q2}
p_a=\frac{\partial {\cal L}}{\partial
\dot{a}}=18\dot{a}aR^{1/2}+\frac{9}{2}a^2\dot{R}R^{-1/2},\end{equation}
\begin{equation}\label{Q3}
p_R=\frac{\partial {\cal L}}{\partial
\dot{R}}=\frac{9}{2}\dot{a}a^2R^{-1/2}.
\end{equation}Therefore, the corresponding Hamiltonian reads
\begin{equation}\label{Q4}
{\cal H}=\frac{2}{9}a^{-2}R^{1/2}p_ap_R-\frac{4}{9}a^{-3}R^{3/2}p_R^2-\frac{1}{2}a^3R^{3/2}.\end{equation}
Although, the classical equations of motion resulting from the Lagrangian (\ref{Q1}) or Hamiltonian (\ref{Q4}) can be solved to give the
corresponding classical cosmology, Hamiltonian (\ref{Q4}) has not the desired form for the construction of the Wheeler-DeWitt equation describing the relevant
quantum cosmology. Furthermore, the Lagrangian (\ref{Q1}) does not exhibit the existence of a cyclic variable corresponded to the Noether symmetry.
To be more precise, we seek a point transformation $(a,R)\rightarrow (u,v)$ on the vector field (\ref{J}) such that in terms of the new variables
$(u,v)$, the Lagrangian includes one cyclic variable. A general discussion of this issue can be found in \cite{9}. Under such point transformation it is easy to show that the vector field (\ref{J}) takes the
form
\begin{equation}\label{Q5}
\tilde{X}=(Xu)\frac{\partial}{\partial u}+(Xv)\frac{\partial}{\partial v}+\frac{d}{dt}(Xu)\frac{\partial}{\partial \dot{u}}+\frac{d}{dt}(Xv)
\frac{\partial}{\partial \dot{v}}.\end{equation}
One can show that if $X$ is a Noether symmetry of the Lagrangian, $\tilde{X}$ has also this property, that is
\begin{equation}\label{Q6}
X{\cal L}=0\Rightarrow \tilde{X}{\cal L}=0.\end{equation}Thus, if we demand
\begin{equation}\label{Q6}
Xu=1,\hspace{.5cm}Xv=0,\end{equation}
we get
\begin{equation}\label{Q7}
\tilde{X}=\frac{\partial}{\partial u}\Rightarrow\tilde{X}{\cal L}=\frac{\partial {\cal L}}{\partial u}=0.\end{equation}
This means that $u$ is a cyclic variable and the dynamics can be reduced. On the other hand, the constant of motion $Q$ becomes
\begin{eqnarray}\label{Q8}
Q=\alpha p_a+\beta p_R=\alpha \frac{\partial {\cal L}}{\partial \dot{a}}+\beta \frac{\partial {\cal L}}{\partial \dot{R}}=\nonumber \\
\alpha \left(\frac{\partial {\cal L}}{\partial u}\frac{\partial u}{\partial \dot{a}}+\frac{\partial {\cal L}}{\partial v}
\frac{\partial v}{\partial \dot{a}}+\frac{\partial {\cal L}}{\partial \dot{u}}\frac{\partial \dot{u}}{\partial \dot{a}}
+\frac{\partial {\cal L}}{\partial \dot{v}}\frac{\partial \dot{v}}{\partial \dot{a}}\right)
+\beta \left(\frac{\partial {\cal L}}{\partial u}\frac{\partial u}{\partial \dot{R}}+\frac{\partial {\cal L}}{\partial v}
\frac{\partial v}{\partial \dot{R}}+\frac{\partial {\cal L}}{\partial \dot{u}}\frac{\partial \dot{u}}{\partial \dot{R}}
+\frac{\partial {\cal L}}{\partial \dot{v}}\frac{\partial \dot{v}}{\partial \dot{R}}\right).
\end{eqnarray}
Since $(a,R)\rightarrow (u,v)$ is a point transformation, we have
\[\frac{\partial u}{\partial \dot{a}}=\frac{\partial v}{\partial \dot{a}}=\frac{\partial u}{\partial \dot{R}}=\frac{\partial v}{\partial \dot{R}}=0,\]
and
\[\frac{\partial \dot{u}}{\partial \dot{a}}=\frac{\partial u}{\partial a},\hspace{.5cm}\frac{\partial \dot{u}}{\partial \dot{R}}=
\frac{\partial u}{\partial R},\hspace{.5cm}\frac{\partial \dot{v}}{\partial \dot{a}}=\frac{\partial v}{\partial a},\hspace{.5cm}\frac{\partial \dot{u}}{\partial \dot{R}}=
\frac{\partial v}{\partial R}.\]Therefore,
\begin{eqnarray}\label{Q9}
Q=\alpha \left(\frac{\partial {\cal L}}{\partial \dot{u}}\frac{\partial u}{\partial a}
+\frac{\partial {\cal L}}{\partial \dot{v}}\frac{\partial v}{\partial a}\right)
+\beta \left(\frac{\partial {\cal L}}{\partial \dot{u}}\frac{\partial u}{\partial R}
+\frac{\partial {\cal L}}{\partial \dot{v}}\frac{\partial v}{\partial R}\right)=\nonumber\\
\left(\alpha \frac{\partial u}{\partial a}+\beta \frac{\partial u}{\partial R}\right)\frac{\partial {\cal L}}{\partial \dot{u}}
+\left(\alpha \frac{\partial v}{\partial a}+\beta \frac{\partial v}{\partial R}\right)\frac{\partial {\cal L}}{\partial \dot{v}}=\nonumber\\
(Xu)\frac{\partial {\cal L}}{\partial \dot{u}}+(Xv)\frac{\partial {\cal L}}{\partial \dot{v}}=\frac{\partial {\cal L}}{\partial \dot{u}}=p_u .
\end{eqnarray}
Thus, as expected the constant of motion which corresponds to the Noether symmetry is nothing but the momentum conjugated to the cyclic variable. To find the explicit form of the above mentioned point transformation we should solve the equations (\ref{Q6}), which give
\begin{equation}\label{Q10}
-\frac{1}{2}a^{-1}\frac{\partial u}{\partial a}+Ra^{-2}\frac{\partial u}{\partial R}=1,\end{equation}
\begin{equation}\label{Q11}
-\frac{1}{2}a^{-1}\frac{\partial v}{\partial a}+Ra^{-2}\frac{\partial v}{\partial R}=0.\end{equation}
These differential equations admit the following general solutions
\begin{equation}\label{Q12}
u(a,R)=-a^2+\phi_1(aR^{1/2}),\hspace{.5cm}v(a,R)=\phi_2(aR^{1/2}),\end{equation}
where $\phi_1$ and $\phi_2$ are two arbitrary functions of $aR^{1/2}$. As is indicated in \cite{9}, "the change of coordinates is not unique and a clever choice is always important." With a glance at the Lagrangian (\ref{Q1}), we choose the functions $\phi_1$ and $\phi_2$ as
\begin{equation}\label{Q13}
\phi_1(aR^{1/2})=(aR^{1/2})^{\mu},\hspace{.5cm}\phi_2(aR^{1/2})=(aR^{1/2})^{\nu},\end{equation}
where $\mu$ and $\nu$ are some constants. With this choice, the Lagrangian (\ref{Q1}) takes the form
\begin{equation}\label{Q14}
{\cal L}=\frac{9}{2}\frac{\mu}{\nu^2}v^{\frac{\mu-2\nu+1}{\nu}}\dot{v}^2-\frac{9}{2\nu}v^{\frac{1-\nu}{\nu}}\dot{u}\dot{v}+\frac{1}{2}v^{3/\nu}.
\end{equation}It is clear from this Lagrangian that $u$ is cyclic and the Noether symmetry is given by $p_u=Q=\mbox{cons.}$. Also, the momenta conjugate to $u$ and $v$ are
\begin{equation}\label{Q15}
p_u=\frac{\partial {\cal L}}{\partial \dot{u}}=-\frac{9}{2\nu}v^{\frac{1-\nu}{\nu}}\dot{v},\hspace{.5cm}
p_v=\frac{\partial {\cal L}}{\partial \dot{v}}=9\frac{\mu}{\nu^2}v^{\frac{\mu-2\nu+1}{\nu}}\dot{v}-\frac{9}{2\nu}v^{\frac{1-\nu}{\nu}}\dot{u},\end{equation}
which give rise to the following Hamiltonian for our dynamical system
\begin{equation}\label{Q16}
{\cal H}=-\frac{10}{9}\mu v^{\frac{\mu-1}{\nu}}p_u^2-\frac{2}{3}\nu v^{\frac{\nu-1}{\nu}}p_u p_v -\frac{1}{2}v^{3/\nu}.\end{equation}
The preliminary set-up for writing the action is now complete. In the next section, we shall focus attention on the study of the quantum cosmology
of the model described above.
\section{Quantization of the model}
Standard cosmological models based on classical general relativity have no convincing
precise answer for the presence of the so-called ''Big-Bang'' singularity. Any hope of dealing
with such singularities would be in vein unless a reliable quantum theory of gravity can
be constructed. In the absence of a full theory of quantum gravity, it would be useful to
describe the quantum states of the universe within the context of quantum cosmology \cite{12}. In this formalism which is
based on the canonical quantization procedure, one first freezes a large number of degrees
of freedom and then quantizes the remaining ones. The quantum state of the universe is
then described by a wave function in the minisuperspace, a function of the 3-geometry
of the model and matter fields presented in the theory, satisfying the Wheeler-DeWitt
equation, that is, ${\cal H}\Psi = 0$, where ${\cal H}$ is the operator form of the
Hamiltonian given by equation (\ref{Q16}) and $\Psi$ is the wave function of the universe. On the other hand, the existence of a Noether symmetry
in the model reduces the dynamics through $p_u=Q$, where its quantum version can be considered as a constraint $p_u \Psi =Q \Psi$. Therefore, the quantum cosmology of our Noether symmetric $f(R)$ model can be described by the following equations
\begin{equation}\label{Q17}
{\cal H}\Psi(u,v)=\left[-\frac{10}{9}\mu v^{\frac{\mu-1}{\nu}}p_u^2-\frac{1}{3}\nu \left(v^r p_v v^s+v^s p_v v^r\right)p_u-\frac{1}{2}v^{3/\nu}\right]\Psi(u,v)=0,\end{equation}
\begin{equation}\label{Q18}
p_u \Psi(u,v)=Q\Psi(u,v),\end{equation}where the parameters $r$ and $s$ satisfy $r+s=\frac{\nu-1}{\nu}$ and denote the ambiguity in the ordering of factors $v$ and $p_v$ in the second
term of (\ref{Q16}). With the replacement $p_u\rightarrow -i\frac{\partial}{\partial u}$ and similarly for
$p_v$ the above equations read
\begin{equation}\label{Q19}
\left[\frac{10}{9}\mu v^{\frac{\mu-1}{\nu}}\frac{\partial^2}{\partial u^2}+\frac{\nu-1}{3}v^{-1/\nu}\frac{\partial}{\partial u}+
\frac{2}{3}\nu v^{\frac{\nu-1}{\nu}}\frac{\partial^2}{\partial u \partial v}-\frac{1}{2}v^{3/\nu}\right]\Psi(u,v)=0,\end{equation}
\begin{equation}\label{Q20}
-i\frac{\partial}{\partial u}\Psi(u,v)=Q\Psi(u,v).\end{equation}The solutions of the above differential equations are separable and may be written in the form $\Psi(u,v)=U(u)V(v)$. Equation (\ref{Q20}) can be immediately integrated leading to a oscillatory behavior for the wave function in
$u$ direction, i.e. in the direction of symmetry, that is
\begin{equation}\label{Q21}
\Psi(u,v)=e^{iQu}V(v).\end{equation}
Substitution this result into relation (\ref{Q19}) yields the following equation for the function $V(v)$
\begin{equation}\label{Q22}
\frac{dV}{dv}=\left[-\frac{5}{3}i\frac{\mu}{\nu}Qv^{\frac{\mu-\nu}{\nu}}-\frac{3}{4\nu Q}iv^{\frac{4-\nu}{\nu}}-\frac{\nu-1}{2\nu}v^{-1}\right]V(v),
\end{equation}with solution
\begin{equation}\label{Q23}
V(v)=v^{\frac{1-\nu}{2\nu}}e^{-i\left(\frac{5}{3}Qv^{\mu/\nu}+\frac{3}{16Q}v^{4/\nu}\right)}.\end{equation}
Thus, the eigenfunctions of the equations (\ref{Q19}) and (\ref{Q20}) can be written as
\begin{equation}\label{Q24}
\Psi_{\mu \nu}(u,v)=v^{\frac{1-\nu}{2\nu}}e^{i\left(Qu-\frac{5}{3}Qv^{\mu/\nu}-\frac{3}{16Q}v^{4/\nu}\right)}.\end{equation}
We may now write the general solutions to
the Wheeler-DeWitt equation as a superposition of the eigenfunctions with shifted Gaussian weight functions
\begin{equation}\label{Q25}
\Psi(u,v)=\int_{-\infty}^{+\infty}\int_{-\infty}^{+\infty}e^{-\varsigma(\mu-\mu_0)^2}e^{-\sigma(\nu-\nu_0)^2}
v^{\frac{1-\nu}{2\nu}}e^{i\left(Qu-\frac{5}{3}Qv^{\mu/\nu}-\frac{3}{16Q}v^{4/\nu}\right)}d\mu d\nu.\end{equation}
We see that the wave function is a superposition of states of the form $e^{iS}$. In semiclassical approximation for quantum gravity \cite{10},
this type of state represents the correlations between classical trajectories and the peaks of the wave function \cite{9}. Inserting $\Psi \sim e^{iS}$ into Wheeler-DeWitt equation, we are led to the Hamilton-Jacobi equation for $S$. Thus, the classical trajectories can be obtained by rewriting the momenta as
derivative of $S$ with respect to the corresponding variables, that is, $p_q=\frac{\partial S}{\partial q}$. Therefore, in semiclassical limit, by
identifying the exponential factor of (\ref{Q24}) with $S$, we can recover the corresponding classical cosmology
\begin{equation}\label{Q26}
S=Qu-\frac{5}{3}v^{\mu/\nu}-\frac{3}{16Q}v^{4/\nu}.\end{equation}
The classical trajectories, which determine the behavior of the scale factor and Ricci scalar are given by
\[p_u=\frac{\partial S}{\partial u},\hspace{.5cm}p_v=\frac{\partial S}{\partial v}.\]
Using the definition of $p_u$ and $p_v$ in (\ref{Q15}), the equations for the classical trajectories become
\begin{equation}\label{Q27}
-\frac{9}{2\nu}v^{\frac{1-\nu}{\nu}}\dot{v}=Q,\end{equation}
\begin{equation}\label{Q28}
9\frac{\mu}{\nu^2}v^{\frac{\mu-2\nu+1}{\nu}}\dot{v}-\frac{9}{2\nu}v^{\frac{1-\nu}{\nu}}\dot{u}=-\frac{5}{3}\frac{\mu}{\nu}Qv^{\frac{\mu-\nu}{\nu}}
-\frac{3}{4Q\nu}v^{\frac{4-\nu}{\nu}}.\end{equation}
Equation (\ref{Q27}) can be easily integrated leading to
\begin{equation}\label{Q29}
v(t)=\left({\cal Q}t-t_0\right)^{\nu},\end{equation}where ${\cal Q}=-\frac{2}{9}Q$ and $t_0$ is an integrating constant. Substituting the above results into equation (\ref{Q28}) yields
\begin{equation}\label{Q30}
u(t)=\frac{1}{3}\left({\cal Q}t-t_0\right)^{\mu}-\frac{1}{108{\cal Q}^2}\left({\cal Q}t-t_0\right)^4+C,\end{equation}
where $C$ is another integrating constant which we can choose it to be zero. Going back to the variables $a$ and $R$, we obtain the corresponding
classical cosmology as
\begin{equation}\label{Q31}
a(t)=\left[\frac{2}{3}\left({\cal Q}t-t_0\right)^{\mu}+\frac{1}{108{\cal Q}^2}\left({\cal Q}t-t_0\right)^4\right]^{1/2},\end{equation}
\begin{equation}\label{Q32}
R(t)=\frac{\left({\cal Q}t-t_0\right)^2}{\frac{2}{3}\left({\cal Q}t-t_0\right)^{\mu}+\frac{1}{108{\cal Q}^2}\left({\cal Q}t-t_0\right)^4}.\end{equation}
For $t\rightarrow t_0/{\cal Q}$, we have an initial singularity in which the scale factor goes to zero while the Ricci scalar has
a large value. On the other hand, in the late time, the universe evolves with a power law expansion ($a(t)\sim t^2$ for $0<\mu \leq 4$ and $a(t) \sim t^{\mu/2}$ for $\mu >4$) and the scalar curvature goes to zero in this limit.

In general, one of the most important features in quantum cosmology is the recovery of classical cosmology from the corresponding quantum model, or in other words, how can the Wheeler-DeWitt wave functions
predict a classical universe. We see that the oscillatory solutions of the form $\Psi \sim e^{iS}$ for Wheeler-DeWitt equation yield the classical solution (\ref{Q31}) which for positive values of $\mu$ can be viewed as an accelerating cosmology. Since the quantum effects in cosmology are important in the very early times of cosmic evolution, we immediately see that, in this limit, the scale factor has the behavior $a(t)\sim t^{\mu/2}$ for $0<\mu\leq 4$ and $a(t)\sim t^2$ for $\mu>4$, which for $\mu>2$ describe a power law inflationary behavior. The universe then undergoes to a late-time accelerating phase, also with a power law expansion behavior. An important question in inflationary models is how much inflation do the model predict or in other words, what is the mechanism through which the universe exits from the inflationary epoch and undergoes into radiation or matter-dominated eras. As is well known, this is largely depend on the behavior of a scalar field with which the universe nucleates \cite{13}. In $f(R)$ gravity models some of such mechanisms describing the transition between different epochs of cosmic evolution are proposed in \cite{14} by introducing a scalar field $\phi$. Indeed, it is easy to verify that the metric $f(R)$ gravity action (\ref{A}) is dynamically equivalent to \cite{TS}
\begin{equation}\label{rev1}
{\cal S}=\int d^4x \sqrt{-g}\left[\phi R-V(\phi)\right],\end{equation}where $\phi=f'(R)$ and $V(\phi)=\phi R(\phi)-f(R(\phi))$. In our $f(R)\sim R^{3/2}$ model we have $R(\phi)\sim \phi^2$ and hence $V(\phi)\sim \phi^3$. The dynamics of this scalar field may show the the possibility
of the transition between different eras of cosmic evolution. But as is indicated in \cite{14} "we should note
that in this case, there is no matter and $f(R)$-terms contribution plays the role of the matter instead of the real matter". Therefore, it is to be noted that our presentation does not claim to clear the role of inflation scenario in a fundamental way because we just study the problem in a special simple model. However, this may reflect realistic scenarios in similar investigations which deal with this problem in a more fundamental way \cite{14}.

\begin{figure}\begin{center}
\epsfig{figure=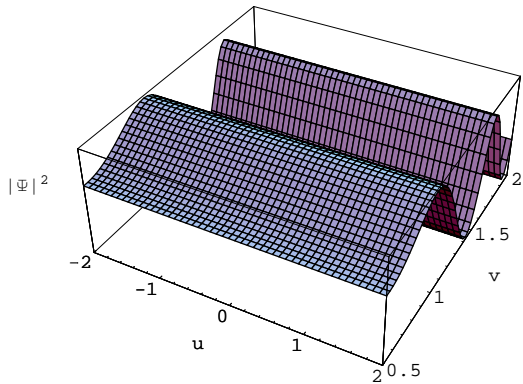,width=7cm} \caption{\footnotesize The square of wave function. The figure is plotted for numerical values $\varsigma =10$, $\sigma =5$, $\mu_0=3$, $\nu_0=0$ and $Q=3$.}\label{fig1}\end{center}
\end{figure}

From equations (\ref{Q29}) and (\ref{Q30}) we see that the classical trajectories obey the relation
\begin{equation}\label{Q33}
u=\frac{1}{3}v^{\mu/\nu}-\frac{1}{108{\cal Q}^2}v^{4/\nu}.\end{equation}As we have mentioned above (see (\ref{Q25})), we are looking for a coherent
wave function with a good asymptotic behavior in the minisuperspace and peaking in the vicinity of the classical loci (\ref{Q33}) in the configuration
space spanned by $\{u,v\}$. It is well known that the general solution of Wheeler-DeWitt equation may be constructed by superposition of its eigenfunctions which
in our problem at hand are labeled by $\mu$ and $\nu$. Therefore, the wavepacket (\ref{Q25}) is what we need. We take the solution as being represented by equation (\ref{Q25}) with the integrals to be truncated at suitable values of $\mu$ and $\nu$ displaying this peak. Figure \ref{fig1} shows the square of wave function for typical values of the parameters where we have taken the integrals from 0 to 2 for $\mu$ and from -15 to 15 for $\nu$. It is seen that almost a good correlation exists between this pattern and the classical trajectories.

\section{Conclusions}
In this letter we have studied a generic $f(R)$ cosmological model by Noether symmetry approach. For the background geometry, we
have considered a flat FRW metric and derived the
general equations of motion in this background. The phase space was
then  constructed by taking the scale factor $a$ and Ricci scalar
$R$ as the independent dynamical variables. The Lagrangian of the
model in the configuration space spanned by $\left\{a,R\right\}$ is
so constructed such that its variation with respect to these
dynamical variables yields the correct field equations. The
existence of Noether symmetry implies that the Lie derivative of
this Lagrangian with respect to the infinitesimal generator of the
desired symmetry vanishes. In \cite{5} we have shown that by applying this condition to the
Lagrangian of the model, one can obtain the explicit form of the
corresponding $f(R)$ function which has led us to the Lagrangian (\ref{Q1}) and Hamiltonian (\ref{Q4}) of the Noether symmetric model. Since the Lagrangian (\ref{Q1}) does not exhibit the existence of a cyclic variable corresponded to the Noether
symmetry, we have provided a point transformation $(a,R)\rightarrow (u,v)$ such that in terms of the new variables
$(u,v)$, the Lagrangian includes one cyclic variable. We have then quantized the model and shown that the corresponding
quantum cosmology and the ensuing Wheeler-DeWitt equation are amenable to exact solutions in terms of a superposition of states of the form $e^{iS}$ due to the existence of Noether symmetry. In semiclassical approximation
for quantum gravity, this type of state represents the correlations between classical trajectories and the peaks of the wave function. Using this
interpretation we have shown that the corresponding classical cosmology results in a power law accelerated expansion for the scale factor of the universe either in its early or late time evolution.

\end{document}